\documentclass{sig-alternate}

\conferenceinfo{EC'10,} {June 7--11, 2010, Cambridge, Massachusetts, USA.}
\CopyrightYear{2010}
\crdata{978-1-60558-822-3/10/06}
\usepackage{xspace}
\usepackage{amsmath}
\usepackage{amssymb}
\usepackage{mathrsfs}
\usepackage{array}
\usepackage{subfigure}
\usepackage{tikz}
\usetikzlibrary{arrows,decorations,decorations.shapes,backgrounds,shapes}

\newtheorem{theorem}{Theorem}[section]
\newtheorem{lemma}[theorem]{Lemma}
\newtheorem{corollary}[theorem]{Corollary}
\newtheorem{conjecture}[theorem]{Conjecture}

\newtheorem{example}[theorem]{Example}

\newcommand{\match}{\textsc{Match}_{\Pi}\xspace}
\newcommand{\mixnmatch}{\textsc{Mix-and-Match}\xspace}

\begin{document}

\title{Mix and Match}

\numberofauthors{4}
\author{
\alignauthor Itai Ashlagi \\
  \affaddr{Harvard Business School} \\
  \affaddr{Boston, MA 02163, USA} \\
  \email{iashlagi@hbs.edu}
\alignauthor Felix Fischer \\
  \affaddr{Harvard SEAS} \\
  \affaddr{Cambridge, MA 02138, USA} \\
	\email{fischerf@seas.harvard.edu}
\and
\alignauthor Ian A. Kash \\
  \affaddr{Harvard SEAS} \\
  \affaddr{Cambridge, MA 02138, USA} \\
	\email{kash@seas.harvard.edu}
\alignauthor Ariel D. Procaccia \\
  \affaddr{Harvard SEAS} \\
  \affaddr{Cambridge, MA 02138, USA} \\
	\email{arielpro@seas.harvard.edu}
}


\maketitle

\begin{abstract}
Consider a matching problem on a graph where disjoint sets of vertices are privately owned by self-interested agents. An edge between a pair of vertices indicates compatibility and allows the vertices to match. We seek a mechanism to maximize the number of matches despite self-interest, with agents that each want to maximize the number of their
own vertices that match. Each agent can choose to hide some of its vertices, and then privately match the hidden vertices with any of its own vertices that go unmatched by the mechanism. A prominent application of this model is to kidney exchange, where agents correspond to hospitals and vertices to donor-patient pairs. Here hospitals may game an exchange by holding back pairs and harm social welfare.

In this paper we seek to design mechanisms that are strategyproof, in the sense that agents cannot benefit from hiding vertices, and approximately maximize efficiency, i.e., produce a matching that is close in cardinality to the maximum cardinality matching. Our main result is the design and analysis of the eponymous \mixnmatch mechanism; we show that this randomized mechanism is strategyproof and provides a 2-approximation.  Lower bounds establish that the mechanism is near optimal.
\end{abstract}

\category{F.2}{Theory of Computation}{Analysis of Algorithms and
Problem Complexity}
\category{J.4}{Computer Applications}{Social and Behavioral
Sciences}[Economics]

\terms{Algorithms, Theory, Economics}

\keywords{Approximate mechanism design without money, Kidney exchange}

\medskip

\section{Introduction}

Treatment for many types of kidney disease relies on transplantation of a kidney.   Since humans have two kidneys and only need one to survive, many patients have a family member or friend willing to donate them a kidney. However, for various reasons, not all potential  donors are compatible with their desired recipient.  A recent innovation is the idea of a kidney exchange, where incompatible donor-patient pairs $u$ and $v$ such that the donor of pair $u$ is compatible with the patient of pair $v$ and the donor of pair $v$ is compatible with the patient of pair $u$ can effectively trade kidneys.

Kidney exchanges have attracted researchers from economics (see, e.g.,~\cite{RothKidneyQJE,RothKidneyJET,RothKidneyAERPP,RothKidneyAER}) and computer science (see, e.g.,~\cite{ABS07,BRM09}).  Much of this research considers the incentives of the donor-patient pairs participating in exchanges.  However, as centralized kidney exchanges are growing, another important decision maker becomes involved: hospitals.  Hospitals act as agents to find matches for their patients.  To do so they may join a group of hospitals that pool their patients and try to find as many matches as possible.  This introduces strategic issues for the hospitals: they may be able to match more of their patients if they selectively omit some of them from the pool and instead match them with other local patients.  While this may be better for the hospital's patients, it is worse for other patients that might have been matched with them.

In this paper we examine an abstract model of exchanges where each agent acts on behalf of a set of clients to find exchanges.  In addition to kidney exchanges this model applies to other settings such as house exchanges (where the agents are real estate agents and the clients are house owners).

Consider a set of agents, each of whom has a set of clients who seek to exchange objects with each other. In this paper we assume that each exchange involves two clients and each client can participate in at most one exchange.  We  assume that each agent only cares about its own clients and therefore seeks to maximize the number of its clients that participate in an exchange.\footnote{Exchanges in our model are abstract and it is irrelevant to our results whether or not monetary transfers between clients are involved, as long as the terms of the exchange are always the same.}

A centralized mechanism to which agents report their clients' information can enable exchanges between clients of different agents. This yields more exchanges and increases social welfare compared to a situation in which each agent operates on its own. In other words, we would like agents to share their individual databases of clients and apply a centralized matching algorithm to the resulting global database.


More formally, the possible exchanges among a set of clients can be represented as an undirected graph in which the vertices represent clients and an undirected edge between clients $u$ and $v$ means that $u$ and $v$ can exchange their objects.
A matching of the graph then corresponds to a set of two-way exchanges. We assume that the set of clients of each agent is private information of that agent, while for any pair of clients it is verifiable by anyone whether an exchange can be conducted between them.
Each agent wishes to maximize the number of its own clients that are matched.

A mechanism receives the graph induced by the agents' reported subsets of clients, and outputs a matching of that graph.  However, since agents need not report all their clients, a second stage takes place in which each agent finds a matching in the graph induced by the set of its clients that have not been matched in the first stage, which includes those clients not reported by the agent, and those reported but not matched by the mechanism.  The utility of the agent then equals the number of its clients that were matched in one of the two stages.

The above model was first studied by Roth et al.~\cite{RSMunpublished} and Ashlagi et al.~\cite{AGR09} in order to deal with incentives of the hospitals. Roth et al.~\cite{RSMunpublished} observed that there are no efficient and strategyproof (SP) mechanisms, where strategyproofness means that it is a dominant strategy for each agent to report all of its clients.  Ashlagi et al.\@ then proceeded to study efficient mechanisms in the Bayesian setting (see Section~\ref{sec:related} for more details).

In this paper we take a fundamentally different, prior-free approach to the nonexistence of efficient and SP mechanisms, by relaxing efficiency rather than strategyproofness.
More specifically, we study the strength of SP mechanisms in terms of the fraction of social welfare they recover.  We say that a mechanism is an \emph{$\alpha$-approximation} mechanism if the cardinality of the maximum cardinality matching is always at most $\alpha$ times the cardinality of the matching returned by the mechanism.\footnote{Since the cardinality of a matching is exactly twice its social welfare, approximating the cardinality of the maximum cardinality matching is equivalent to approximating the maximum social welfare.} Our goal is to design mechanisms that are SP and at the same time provide a good approximation ratio.

Since in most countries it is both illegal and considered immoral to make payments in return for organs, we are interested in mechanisms without payments.  This goal is compatible with the agenda of \emph{approximate mechanism design without money}~\cite{PT09}. 

\par\medskip
\noindent\textbf{Our results.}
In Section~\ref{sec:lb} we establish some lower bounds. Our starting point is an example, due to Roth et al.~\cite{RSMunpublished} (see also Ashlagi et al.~\cite{AGR09}), which implies that a mechanism that always returns an optimal matching cannot be strategyproof. We refine their result by observing that no deterministic SP mechanism can provide an approximation ratio better than two, and no randomized SP mechanism can provide an approximation ratio better than $4/3$ (Theorem~\ref{thm:lb}).

In Section~\ref{sec:det} we introduce a mechanism, termed $\match$, that is parameterized by a bipartition $\Pi=(\Pi_1,\Pi_2)$ of the agents. Roughly speaking, given a graph the mechanism returns a matching that has maximum cardinality among all the matchings that (i) contain no edges between the vertex sets of two agents on the same side of the bipartition, and (ii) are a maximum cardinality matching when restricted to the vertex set of each individual agent. Our main technical result is the following theorem.

\par\medskip
\textsc{Theorem~\ref{thm:sp}.}
\emph{For any number of agents, and for any bipartition $\Pi$ of the set of agents, $\match$ is SP.}
\medskip

The main idea behind the proof of this theorem rather subtle.  It relies on the fact that if one takes the union of the two matchings produced by the mechanism before and after an agent hides some of its vertices, then this union cannot contains a cycle that visits the vertex sets of an odd number of agents. This property holds because the mechanism does not match edges between vertex sets of agents on the same side of the bipartition.

We further show that $\match$ can be executed in polynomial time. Unfortunately, for any deterministically fixed bipartition $\Pi$, $\match$ does not generally provide a bounded approximation ratio. We however observe that $\match$ yields a $2$-approximation in the two agent case when used with the obvious bipartition that places the two agents on opposite sides (Corollary~\ref{cor}). This mechanism is in fact the optimal deterministic SP mechanism when there are two agents, since the deterministic lower bound of~$2$ holds even in this case.

In Section~\ref{sec:rand} we consider randomized mechanisms, and leverage Theorem~\ref{thm:sp} to establish a strong randomized upper bound.  We introduce a mechanism, termed \mixnmatch, which consists of two steps: the mechanism first \emph{mixes} the agents by choosing a random bipartition $\Pi$, then \emph{matches} the vertices by applying $\match$. The following theorem is conceptually the main result of the paper.

\par\medskip
\textsc{Theorem~\ref{thm:approx}.}
\emph{For any number of agents, \mixnmatch is SP and provides a $2$-approximation with respect to social welfare.}
\medskip

Strategyproofness of the mechanism follows from Theorem~\ref{thm:sp} in a straightforward way, but its approximation guarantees are nontrivial.

\par\medskip
\noindent\textbf{Open problems and future work.}
There are several gaps between our upper and lower bounds. The most enigmatic gap concerns deterministic mechanisms when the number of agents is at least three. While Theorem~\ref{thm:lb} provides a deterministic lower bound of $2$, we were unable to design a deterministic SP mechanism with a constant approximation ratio, and indeed we conjecture that such a mechanism does not exist (Conjecture~\ref{conj:det_lb}).

With respect to randomized mechanisms, there is a small gap between the lower bound of $4/3$ and the upper bound of $2$ provided by \mixnmatch.  In Section~\ref{sec:rand}, we present a mechanism, termed \textsc{Flip-and-Match}, for the case of two agents, which clearly provides a $(4/3)$-approximation.  We conjecture that this mechanism is also SP (Conjecture~\ref{conj:rand_ub}), and discuss this conjecture in Appendix~\ref{app:rand_ub}. Our unfounded guess is that the randomized lower bound for more than two agents is $2$.

There also exist a number of possible extensions to our work, of which we briefly point out a few. As we assume that agents wish to maximize the number of their clients being matched, a natural and realistic extension would be to incorporate weights into the model. For example, a client may value two feasible exchanges differently, or some clients may be more important than others.  Another direction would be to allow exchanges of length greater than two; in this case one should consider directed graphs and look for sets of disjoint cycles that cover many vertices.  Finally, we could ask for the stronger requirement of group-strategyproofness, which requires that no group of agents would want to deviate in a coordinated fashion.  A related approach would be to consider solution concepts like the core, which ensure that a group of hospitals would not want to leave and form a smaller pool.  In the case of kidney exchanges this however seems rather unlikely (at least in our simple model), because hospitals would presumably not want to leave a pool to help a few patients in the current match at the cost of not having access to the pool for all their future patients.

\section{Related Work}
\label{sec:related}

\par\medskip
\noindent\textbf{Work on incentives in kidney exchange.}
The incentives of incompatible donor-patients pairs have been studied for quite some time~\cite{RothKidneyQJE,RothKidneyJET}.  However, as centralized kidney exchange clearinghouses grow, hospitals' incentives become a real issue. For example, reports from the Alliance for Paired Donation indicate that hospitals do not commit to assigned exchanges and perform internal exchanges instead. 

To this end, Roth et al.~\cite{RSMunpublished} introduced a model of kidney exchanges in which hospitals are agents.  They show that no individually rational (IR) and efficient mechanism can be SP, where individual rationality requires that a mechanism matches for each agent at least the number of clients that it can match on its own with respect to its reported set of clients. In our model individual rationality is a special case of strategyproofness. Ashlagi et al.~\cite{AGR09} demonstrate that there exists an $\epsilon$-Bayesian incentive compatible IR mechanism that is also efficient. Furthermore, they show that exchanges of size at most four are sufficient to reach efficiency for large enough graphs. The priors in their Bayesian setting stem from data-driven parameters like the structure and frequency of blood types; no such information is required in our prior-free setting.

\par\medskip
\noindent\textbf{Work on approximate mechanism design without money.}
Procaccia and Tennenholtz~\cite{PT09} recently introduced the notion of approximate mechanism design without money, which was already implicit in earlier work on truthful learning~\cite{DFP10}. The starting point is the large body of work on algorithmic mechanism design, the study of truthful approximation mechanisms for game-theoretic versions of optimization problems (see, e.g.,~\cite{NR01,LOS02,LS05,CKV09,DDDR08}).
The mechanisms in this area typically allow the use of monetary payments to align agents' incentives. Procaccia and Tennenholtz argue that the use of money is infeasible in many settings due to ethical, legal, or practical considerations.  In such cases it is more desirable to design SP mechanisms that are only approximately efficient but do not utilize payments. This approach is particularly interesting in the context of computationally tractable optimization problems: while there is no need to resort to approximate solutions for computational reasons, they might be used to achieve strategyproofness when the optimal solution is not SP.

Procaccia and Tennenholtz~\cite{PT09} specifically study a facility location problem on the real line. Some of their results were recently improved by Lu et al.~\cite{LWZ09}, and extended to graphs by Alon et al.~\cite{AFPT09}. Other domains to which the above principle was recently applied include the selection of vertices in a directed graph~\cite{AFPT09b}, the allocation of items~\cite{GC10}, and classification~\cite{MPR08b,MPR09}.

Most recently, Dughmi and Ghosh~\cite{DG10} studied approximate mechanism design without money in the context of the generalized assignment problem. Their setting consists of a bipartite graph with jobs on one side and machines on the other, where machines have capacities and edges have values and sizes. The agents in their setting are the jobs (which in our setting would be the clients rather than the agents), and edges incident to a job are private information of that job. Dughmi and Ghosh in fact briefly consider maximum matching as a first special case of their model, but their motivation, setting and results are all fundamentally different from ours. In particular, in the context of maximum unweighted matching their model easily admits an SP \emph{optimal} mechanism, whereas this problem is quite intricate in our model and forms the topic of this entire paper. 


\section{Preliminaries}
\label{sec:prem}

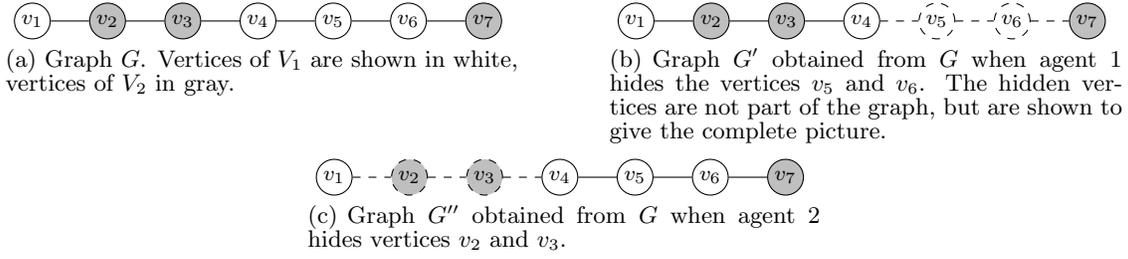
\begin{figure*}
\centering
\subfigure[Graph $G$. Vertices of $V_1$ are shown in white, vertices of $V_2$ in gray.]
{
\begin{tikzpicture}
\tikzstyle{graydot}=[circle,draw=black,fill=lightgray,thin,
inner sep=1.5pt,minimum size=1.5mm]
\tikzstyle{whitedot}=[circle,draw=black,fill=white,thin,
inner sep=1.5pt,minimum size=1.5mm]
\tikzstyle{dasheddot}=[circle,dashed,draw=black,fill=white,thin,
inner sep1.5pt,minimum size=1.5mm]

\node (v1) at (0,0) [whitedot] {\small{$v_1$}};
\node (v2) at (1,0) [graydot] {\small{$v_2$}};
\node (v3) at (2,0) [graydot] {\small{$v_3$}};
\node (v4) at (3,0) [whitedot] {\small{$v_4$}};
\node (v5) at (4,0) [whitedot] {\small{$v_5$}};
\node (v6) at (5,0) [whitedot] {\small{$v_6$}};
\node (v7) at (6,0) [graydot] {\small{$v_7$}};

\draw (v1.east) -- (v2.west);
\draw (v2.east) -- (v3.west);
\draw (v3.east) -- (v4.west);
\draw (v4.east) -- (v5.west);
\draw (v5.east) -- (v6.west);
\draw (v6.east) -- (v7.west);

\end{tikzpicture}
\label{fig:lb_a}
}
\hspace{1cm}
\subfigure[Graph $G'$ obtained from $G$ when agent $1$ hides the vertices $v_5$ and $v_6$. The hidden vertices are not part of the graph, but are shown to give the complete picture.]
{
\begin{tikzpicture}
\tikzstyle{graydot}=[circle,draw=black,fill=lightgray,thin,
inner sep=1.5pt,minimum size=1.5mm]
\tikzstyle{whitedot}=[circle,draw=black,fill=white,thin,
inner sep=1.5pt,minimum size=1.5mm]
\tikzstyle{dasheddot}=[circle,dashed,draw=black,fill=white,thin,
inner sep=1.5pt,minimum size=1.5mm]

\node (v1) at (0,0) [whitedot] {\small{$v_1$}};
\node (v2) at (1,0) [graydot] {\small{$v_2$}};
\node (v3) at (2,0) [graydot] {\small{$v_3$}};
\node (v4) at (3,0) [whitedot] {\small{$v_4$}};
\node (v5) at (4,0) [dasheddot] {\small{$v_5$}};
\node (v6) at (5,0) [dasheddot] {\small{$v_6$}};
\node (v7) at (6,0) [graydot] {\small{$v_7$}};

\draw (v1.east) -- (v2.west);
\draw (v2.east) -- (v3.west);
\draw (v3.east) -- (v4.west);
\draw [dashed] (v4.east) -- (v5.west);
\draw [dashed] (v5.east) -- (v6.west);
\draw [dashed] (v6.east) -- (v7.west);

\end{tikzpicture}
\label{fig:lb_b}
}
\hspace{1cm}
\subfigure[Graph $G''$ obtained from $G$ when agent $2$ hides vertices $v_2$ and $v_3$.]
{
\begin{tikzpicture}
\tikzstyle{graydot}=[circle,draw=black,fill=lightgray,thin,
inner sep=1.5pt,minimum size=1.5mm]
\tikzstyle{whitedot}=[circle,draw=black,fill=white,thin,
inner sep=1.5pt,minimum size=1.5mm]
\tikzstyle{dasheddot}=[circle,dashed,draw=black,fill=white,thin,
inner sep=1.5pt,minimum size=1.5mm]
\tikzstyle{dashedgraydot}=[circle,dashed,draw=black,fill=lightgray,thin,
inner sep=1.5pt,minimum size=1.5mm]

\node (v1) at (0,0) [whitedot] {\small{$v_1$}};
\node (v2) at (1,0) [dashedgraydot] {\small{$v_2$}};
\node (v3) at (2,0) [dashedgraydot] {\small{$v_3$}};
\node (v4) at (3,0) [whitedot] {\small{$v_4$}};
\node (v5) at (4,0) [whitedot] {\small{$v_5$}};
\node (v6) at (5,0) [whitedot] {\small{$v_6$}};
\node (v7) at (6,0) [graydot] {\small{$v_7$}};

\draw [dashed] (v1.east) -- (v2.west);
\draw [dashed] (v2.east) -- (v3.west);
\draw [dashed] (v3.east) -- (v4.west);
\draw (v4.east) -- (v5.west);
\draw (v5.east) -- (v6.west);
\draw (v6.east) -- (v7.west);

\end{tikzpicture}
\label{fig:lb_c}
}
\caption{Construction used in the proof of Theorem~\ref{thm:lb}.}
\label{fig:lb}
\end{figure*}

Let $N=\{1,\ldots,n\}$ be a set of agents. For each $i\in N$, let $V_i$ be a set of private vertices of agent~$i$. Let $G=(V,E)$ with $V=\bigcup_{i\in N}{V_i}$ be an undirected labeled graph, that is, each vertex is labeled by its agent. We slightly abuse terminology by simply referring to such labeled graphs as ``graphs.''

A \emph{matching} $M\subseteq E$ on $G$ is a subset of edges such that each vertex is incident to at most one edge of $M$. For $i,j\in N$ we denote
$$
M_{ij} = \{(u,v)\in M:\ u\in V_i\ \wedge\ v\in V_j\}.
$$
Given $i\in N$, we refer to edges in $M_{ii}$ as \emph{internal edges} and to edges in $M_{ij}$, where $j\in N\setminus\{i\}$, as \emph{external edges}.

Given a graph $G$ and a matching $M$ on $G$, the utility of agent $i$ for this matching is
$$
u_i(M) = |\{u\in V_i:\ \exists v\in V s.t.\ (u,v)\in M\}|,
$$
that is, it is equal to the number of vertices of $V_i$ that are matched under $M$.

We now turn to the definition of a mechanism, without being too formal. For a fixed number $n$ of agents, a \emph{deterministic mechanism} is a function that maps any (labeled) graph for $n$ agents to a matchings of this graph. A \emph{randomized mechanism} maps any graph to a probability distribution over matchings, that is, it can select a matching randomly. For conciseness, we treat deterministic mechanisms as a special case of randomized mechanisms in the rest of this section.

For a randomized mechanism $f$ and a (possibly random) graph $G$, define
$$
u_i(f(G)) = \mathbb{E}_{M \sim f(G)}[u_i(M)],
$$
where the expectation is taken over the distribution on matchings returned by the mechanism. In other words, the utility of an agent simply equals the expected number of its vertices being matched.

We are concerned with situations where an agent ``hides'' a subset of its vertices and then internally matches them among themselves or with vertices not matched by the mechanism. To make this formal we need some notation. We however feel that the idea is rather intuitive, and will avoid the rather cumbersome formalism in the rest of the paper. For any subset $V'\subseteq V$, let $G[V']$ be the subgraph of $G$ induced by $V'$. For a graph $G$, an agent $i\in N$, and a matching $M$, let $X_i(M)$ be the set of vertices in $V_i$ that are not matched in $M$; if $M$ is chosen randomly, then $X_i(M)$ is a random variable. Furthermore, let $f^*$ be a mechanism that maps each graph $G$ to a maximum cardinality matching of $G$. We say that a mechanism $f$ is \emph{strategyproof (SP)} if for every graph $G=(V,E)$ with $V=\bigcup_{i\in N}V_i$, for every $i\in N$, and for every $V_i'\subseteq V_i$ it holds that
\begin{align*}
u_i(f(G))&\geq u_i(f(G[V\setminus V_i']))\\
&+u_i(f^*(G[V'_i\cup X_i(f(G[V\setminus V_i']))])).
\end{align*}
In other words, a mechanism is SP if an agent can never benefit by hiding some of his vertices. The agent's utility after hiding a subset $V_i'$ of its vertices equals the (expected) number of its vertices that the mechanism matches given the subgraph induced by all vertices but those in $V_i'$, plus the (expected) number of vertices in a maximum cardinality matching of the subgraph induced by $V_i'$ and the vertices not matched by the mechanism.
In our model, individual rationality (IR) requires that an agent cannot benefit from the special case when $V_i'=V_i$, and is therefore implied by strategyproofness.

We are interested in mechanisms that, while being SP, produce matchings that maximize social welfare, i.e., the sum of agent utilities.  For any matching $M$, $\sum_{i\in N} u_i(M) = 2|M|$, so what we are looking for are matchings that are as large as possible. We say that a randomized mechanism $f$ provides an $\alpha$-approximation if for every graph $G$,
\begin{equation}
\label{eq:approx}
\frac{|f^*(G)|}{\mathbb{E}[|f(G)|]}\leq \alpha,
\end{equation}
where once again $f^*(G)$ is a maximum cardinality matching of $G$. For deterministic mechanisms, the expectation in~\eqref{eq:approx} can simply be dropped.

\section{Lower bounds}
\label{sec:lb}

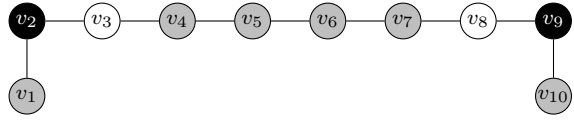
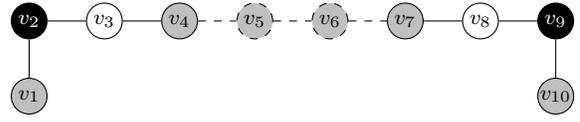
\begin{figure*}
\centering
\subfigure[The original graph $G$, where the vertices of $V_1$ are white, the vertices of $V_2$ are gray, and the vertices of $V_3$ are black.]
{
\begin{tikzpicture}
\tikzstyle{graydot}=[circle,draw=black,fill=lightgray,thin,
inner sep=1.5pt,minimum size=1.5mm]
\tikzstyle{blackdot}=[circle,draw=black,fill=black,thin,
inner sep=1.5pt,minimum size=1.5mm]
\tikzstyle{whitedot}=[circle,draw=black,fill=white,thin,
inner sep=1.5pt,minimum size=1.5mm]
\tikzstyle{graydot2}=[circle,draw=black,fill=lightgray,thin,
inner sep=0.3,minimum size=1.5mm]
\tikzstyle{dasheddot}=[circle,dashed,draw=black,fill=white,thin,
inner sep1.5pt,minimum size=1.5mm]

\node (v1) at (1,-1) [graydot] {\small{$v_1$}};
\node (v2) at (1,0) [blackdot] {\color{white}\small{$v_2$}};
\node (v3) at (2,0) [whitedot] {\small{$v_3$}};
\node (v4) at (3,0) [graydot] {\small{$v_4$}};
\node (v5) at (4,0) [graydot] {\small{$v_5$}};
\node (v6) at (5,0) [graydot] {\small{$v_6$}};
\node (v7) at (6,0) [graydot] {\small{$v_7$}};
\node (v8) at (7,0) [whitedot] {\small{$v_8$}};
\node (v9) at (8,0) [blackdot] {\color{white}\small{$v_9$}};
\node (v10) at (8,-1) [graydot2] {\small{$v_{10}$}};

\draw (v1.north) -- (v2.south);
\draw (v2.east) -- (v3.west);
\draw (v3.east) -- (v4.west);
\draw (v4.east) -- (v5.west);
\draw (v5.east) -- (v6.west);
\draw (v6.east) -- (v7.west);
\draw (v7.east) -- (v8.west);
\draw (v8.east) -- (v9.west);
\draw (v9.south) -- (v10.north);

\end{tikzpicture}
\label{fig:counter_a}
}
\hspace{1cm}
\subfigure[The graph $G'$, agent $2$ hides vertices $v_5$ and $v_6$.]
{
\begin{tikzpicture}
\tikzstyle{graydot}=[circle,draw=black,fill=lightgray,thin,
inner sep=1.5pt,minimum size=1.5mm]
\tikzstyle{blackdot}=[circle,draw=black,fill=black,thin,
inner sep=1.5pt,minimum size=1.5mm]
\tikzstyle{whitedot}=[circle,draw=black,fill=white,thin,
inner sep=1.5pt,minimum size=1.5mm]
\tikzstyle{graydot2}=[circle,draw=black,fill=lightgray,thin,
inner sep=0.3,minimum size=1.5mm]
\tikzstyle{dasheddot}=[circle,dashed,draw=black,fill=lightgray,thin,
inner sep=1.5pt,minimum size=1.5mm]

\node (v1) at (1,-1) [graydot] {\small{$v_1$}};
\node (v2) at (1,0) [blackdot] {\color{white}\small{$v_2$}};
\node (v3) at (2,0) [whitedot] {\small{$v_3$}};
\node (v4) at (3,0) [graydot] {\small{$v_4$}};
\node (v5) at (4,0) [dasheddot] {\small{$v_5$}};
\node (v6) at (5,0) [dasheddot] {\small{$v_6$}};
\node (v7) at (6,0) [graydot] {\small{$v_7$}};
\node (v8) at (7,0) [whitedot] {\small{$v_8$}};
\node (v9) at (8,0) [blackdot] {\color{white}\small{$v_9$}};
\node (v10) at (8,-1) [graydot2] {\small{$v_{10}$}};

\draw (v1.north) -- (v2.south);
\draw (v2.east) -- (v3.west);
\draw (v3.east) -- (v4.west);
\draw [dashed] (v4.east) -- (v5.west);
\draw [dashed] (v5.east) -- (v6.west);
\draw [dashed] (v6.east) -- (v7.west);
\draw (v7.east) -- (v8.west);
\draw (v8.east) -- (v9.west);
\draw (v9.south) -- (v10.north);

\end{tikzpicture}
\label{fig:counter_b}
}
\caption{The na\"ive $3$-agent mechanism is not SP.}
\label{fig:counter}
\end{figure*}

It may not be immediately apparent that the optimal mechanism is not SP. Given a graph, the optimal mechanism simply returns a maximum cardinality matching (while employing a consistent tie-breaking rule to decide between different maximum cardinality matchings).

To see how this can fail to be SP, consider the graph $G$ in Figure~\ref{fig:lb_a}. This graph has an odd number of vertices, so every matching leaves some vertex unmatched. However, each agent has a pair of vertices such that removing these vertices from the graph results in a graph with a unique maximum cardinality matching in which all of that agent's vertices are matched (Figures~\ref{fig:lb_b} and~\ref{fig:lb_c}). Thus, one of the agents must have an unmatched vertex in $G$, and this agent can hide two of his vertices to increase his utility. This simple example, which is due to Roth et al.~\cite{RSMunpublished}, can be used to derive lower bounds that will later turn out to be, at least in one case, tight.
\begin{theorem}
\label{thm:lb}
If there are at least two agents,
\begin{enumerate}
\item no deterministic SP mechanism can provide an $\alpha$-ap\-proximation with respect to social welfare for $\alpha<2$, and
\item no randomized SP mechanism can provide an $\alpha$-ap\-proximation with respect to social welfare for $\alpha<4/3$.
\end{enumerate}
\end{theorem}
\begin{proof}
For the first part of the theorem, we consider the case where $N=\{1,2\}$; the proof can easily be extended to the case where $n>2$ by adding agents with vertices that are not incident to any edges. Let $f$ be a deterministic mechanism, and consider the graph $G$ given in Figure~\ref{fig:lb_a}. Since $G$ has an odd number of vertices, it does not have a perfect matching, so $f(G)$ must leave some $v\in V_1$ or some $v\in V_2$ unmatched. Thus, either $u_1(f(G))\leq 3$ or $u_2(f(G))\leq 2$.

We first deal with the case where $u_1(f(G))\leq 3$. Consider the graph $G'$ that is obtained when agent $1$ hides vertices $v_5$ and $v_6$ (see Figure~\ref{fig:lb_b}). The unique maximum cardinality matching of this graph is $\{(v_1,v_2),(v_3,v_4)\}$, a matching of cardinality $2$. However, agent $1$ could internally match the pair $(v_5,v_6)$ and obtain a utility of $4$, contradicting strategyproofness. Therefore, $f(G')$ must have cardinality at most $1$, meaning that its approximation ratio on $G'$ cannot be smaller than $2$.

The case where $u_2(f(G))\leq 2$ can be handled similarly. Consider the graph $G''$ obtained when agent $2$ hides vertices $v_2$ and $v_3$ (see Figure~\ref{fig:lb_c}). Once again there is a unique maximum matching of cardinality $2$, but $f$ cannot return this matching since it would yield a utility of $3$ to agent $2$, in contradiction to strategyproofness. As before the mechanism is forced to select a matching of cardinality at most~$1$.

The second part of the theorem can be derived using the same construction. Let $f$ be a randomized SP mechanism. Since $G$ does not have a perfect matching, it must be the case that $f(G)$ either does not match some vertex of $V_1$ with probability at least $1/2$, or it does not match some vertex of $V_2$ with probability at least $1/2$, that is, either $u_1(f(G))\leq 7/2$ or $u_2(f(G))\leq 5/2$.

We now proceed as before. If $u_1(f(G))\leq 7/2$, we consider the graph $G'$; by strategyproofness $f$ can only match $3/2$ pairs in expectation, but the optimum is $2$. If $u_2(f(G))\leq 5/2$, we use the graph $G''$ to show that $f$ can only match $3/2$ pairs in expectation, while the optimum is $2$.
\end{proof}

\section{Deterministic Mechanisms}
\label{sec:det}

Let us now consider deterministic mechanisms.  We begin this section by designing a deterministic mechanism that is SP for any number of agents, but may not provide a bounded approximation ratio. We then leverage this mechanism to obtain an optimal deterministic SP mechanism for two agents. The more powerful application of our deterministic mechanism will only appear in the next section, when we discuss randomized mechanisms.

Let us first address the issue of designing SP deterministic mechanisms without worrying, for now, about approximate efficiency or computational tractability. Consider the following mechanism for two agents. Given a graph $G$, the mechanism computes the set of all matchings on $G$ that have maximum cardinality on $V_1$ and $V_2$, and among these selects a matching with maximum overall cardinality. Since every matching that this mechanism considers has maximum cardinality on $V_1$ and $V_2$, it clearly is individually rational.  We will show momentarily that it is also SP.

But let us first consider what this mechanism does when applied to the graph of Figure~\ref{fig:lb_a}. Any matching that is a maximum cardinality matching on $V_2$ would have to match $(v_2,v_3)$, and there are two maximum cardinality matchings on $V_1$: one can either match $(v_4,v_5)$ or $(v_5,v_6)$. If we match $(v_5,v_6)$, no additional edges can be added. Hence, the unique matching of cardinality $3$ that maximizes the number of internal edges is $\{(v_2,v_3),(v_4,v_5),(v_6,v_7)\}$. The only unmatched vertex in this matching is~$v_1$. With the proof of Theorem~\ref{thm:lb} in mind, let us verify that agent $1$ cannot benefit by hiding $v_5$ and $v_6$. Given the graph $G'$ in Figure~\ref{fig:lb_b}, the mechanism would simply return the matching $(v_2,v_3)$, since this is the unique matching that is a maximum cardinality matching on $V_2$.

The two-agent mechanism suggested above seems promising from the perspective of strategyproofness. Let us extend it to an $n$-agent mechanism in the natural way, and consider the mechanism that selects a matching of maximum cardinality among the matchings that have maximum cardinality on each $V_i$, $i=1,\ldots,n$. In addition, let us break ties \emph{serially}: among all the matchings that meet the above criteria, we select a matching that maximizes the utility of agent~$1$; if there are several such matchings, we choose one that maximizes the utility of agent $2$, and so on.

Interestingly enough, this $n$-agent mechanism is not SP, even when $n=3$. Consider the graph $G$ given in Figure~\ref{fig:counter_a}. Any matching that has maximum cardinality on $V_2$ must match $(v_4,v_5)$ and $(v_6,v_7)$; by the tie-breaking rule the mechanism then returns the matching $\{(v_2,v_3),(v_4,v_5),(v_6,v_7),(v_8,v_9)\}$. When agent $2$ hides $v_5$ and $v_6$ we obtain the graph $G'$ given in Figure~\ref{fig:counter_b}. On this graph the mechanism returns a perfect matching $\{(v_1,v_2),(v_3,v_4),(v_7,v_8),(v_9,v_{10})\}$. After internally matching $(v_5,v_6)$ agent $1$ gains two additional matched vertices compared to the matching on $G$. Clearly this example can be modified to work if ties are broken in a different order.

The deeper reason why the above mechanism fails to be strategyproof is rather subtle, and has to do with the following observation. If one takes the union of the matchings generated on the graphs of Figures~\ref{fig:counter_a} and \ref{fig:counter_b}, and contracts each $V_i$ to one vertex, one obtains an odd length cycle between $V_1$, $V_2$, and $V_3$, as the matching on $G$ has an edge between $V_1$ and $V_3$, and the matching on $G'$ has edges between $V_1$ and $V_2$, and $V_2$ and $V_3$. We proceed to refine the above mechanism in order to avoid such odd cycles; this turns out to be sufficient to guarantee strategyproofness. The following is in fact a family of mechanisms, parameterized by a fixed bipartition $\Pi=(\Pi_1,\Pi_2)$ of the set of agents.

\par\medskip
\noindent$\match$
\begin{enumerate}
\item Given a graph $G$, consider all the matchings that have maximum cardinality on each $V_i$ and do not have any edges between $V_i$ and $V_j$ when $i,j\in\Pi_l$ for some $l\in\{1,2\}$, i.e., those that maximize the number of internal edges and do not have any edges between sets on the same side of the bipartition.
\item Among these matchings select one of maximum cardinality, breaking ties serially in favor of agents in $\Pi_1$ and then agents in $\Pi_2$.
\end{enumerate}

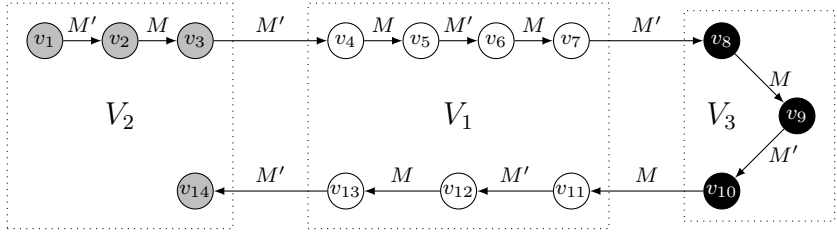
\begin{figure*}
\centering
\begin{tikzpicture}
\tikzstyle{graydot}=[circle,draw=black,fill=lightgray,thin,
inner sep=1.5pt,minimum size=1.5mm]
\tikzstyle{blackdot}=[circle,draw=black,fill=black,thin,
inner sep=1.5pt,minimum size=1.5mm]
\tikzstyle{whitedot}=[circle,draw=black,fill=white,thin,
inner sep=1.5pt,minimum size=1.5mm]
\tikzstyle{graydot2}=[circle,draw=black,fill=lightgray,thin,
inner sep=0.3,minimum size=1.5mm]
\tikzstyle{whitedot2}=[circle,draw=black,fill=white,thin,
inner sep=0.3,minimum size=1.5mm]
\tikzstyle{blackdot2}=[circle,draw=black,fill=black,thin,
inner sep=0.3,minimum size=1.5mm]

\node (v1) at (-1,0) [graydot] {\small{$v_1$}};
\node (v2) at (0,0) [graydot] {\small{$v_2$}};
\node (v3) at (1,0) [graydot] {\small{$v_3$}};
\node (v4) at (3,0) [whitedot] {\small{$v_4$}};
\node (v5) at (4,0) [whitedot] {\small{$v_5$}};
\node (v6) at (5,0) [whitedot] {\small{$v_6$}};
\node (v7) at (6,0) [whitedot] {\small{$v_7$}};
\node (v8) at (8,0) [blackdot] {\color{white}\small{$v_8$}};
\node (v9) at (9,-1) [blackdot] {\color{white}\small{$v_9$}};
\node (v10) at (8,-2) [blackdot2] {\color{white}\small{$v_{10}$}};
\node (v11) at (6,-2) [whitedot2] {\small{$v_{11}$}};
\node (v12) at (4.5,-2) [whitedot2] {\small{$v_{12}$}};
\node (v13) at (3,-2) [whitedot2] {\small{$v_{13}$}};
\node (v14) at (1,-2) [graydot2] {\small{$v_{14}$}};

\path [-latex] (v1) edge node [above] {\small{$M'$}} (v2);
\path [-latex] (v2) edge node [above] {\small{$M$}} (v3);
\path [-latex] (v3) edge node [above] {\small{$M'$}} (v4);
\path [-latex] (v4) edge node [above] {\small{$M$}} (v5);
\path [-latex] (v5) edge node [above] {\small{$M'$}} (v6);
\path [-latex] (v6) edge node [above] {\small{$M$}} (v7);
\path [-latex] (v7) edge node [above] {\small{$M'$}} (v8);
\path [-latex] (v8) edge node [right] {\small{$M$}} (v9);
\path [-latex] (v9) edge node [right] {\small{$M'$}} (v10);
\path [-latex] (v10) edge node [above] {\small{$M$}} (v11);
\path [-latex] (v11) edge node [above] {\small{$M'$}} (v12);
\path [-latex] (v12) edge node [above] {\small{$M$}} (v13);
\path [-latex] (v13) edge node [above] {\small{$M'$}} (v14);

\draw[dotted] (-1.5,0.5) rectangle (1.5,-2.5);
\draw[dotted] (2.5,0.5) rectangle (6.5,-2.5);
\draw[dotted] (7.5,0.4) rectangle (9.5,-2.4);

\node at (0,-1) {\large{$V_2$}};
\node at (4.5,-1) {\large{$V_1$}};
\node at (8,-1) {\large{$V_3$}};

\end{tikzpicture}
\caption{Illustration of Case 1 of the proof of Theorem~\ref{thm:sp}, with $i=1$ as the manipulator, and $\Pi=(\{1\},\{2,3\})$. $M\Delta M'$ is shown as a single directed path with alternating edges of $M$ and $M'$. It holds that $3=|M_{11}|> |M'_{11}|=2$. Every subpath inside $V_2$ and $V_3$ has even length (those from $v_1$ to $v_3$ and from $v_8$ to $v_{10}$), but subpaths inside $V_1$ may not have (like that from $v_4$ and $v_7$). The subpath of $M\Delta M'\setminus (M_{11}\cup M'_{11})$ from $v_1$ to $v_4$ enters $V_1$ but does not exit it, while the subpath from $v_{13}$ to $v_{14}$ exits $V_1$ but does not enter it. This example satisfies~\eqref{eq:diff} with equality.}
\label{fig:partition}
\end{figure*}

By letting $N=\{1,2\}$, $\Pi_1=\{1\}$, and $\Pi_2=\{2\}$, we obtain the two-agent mechanism described above.  The na\"ive generalization of this mechanism to three agents, on the other hand, is not an instance of $\match$: for the example of Figure~\ref{fig:counter} showing that the mechanism is not SP, the sets $M_{12}$, $M_{13}$, and $M_{23}$ are all non-empty.
We proceed to show that $\match$ is SP for any bipartition of the set of agents.
\begin{theorem}
\label{thm:sp}
For any number of agents, and for any bipartition $\Pi$ of the set of agents, $\match$ is SP.
\end{theorem}
\begin{proof}
Fix some bipartition $\Pi=(\Pi_1,\Pi_2)$ of $N$. Consider a graph $G$, and let $M=\match(G)$. Assume that agent $i\in N$ hides a subset of vertices, inducing a subgraph $G'$, and let $M'$ be the matching that results from applying the mechanism to $G'$, along with the internal matching of agent $1$ on its hidden and unmatched vertices, that is,
$$M'=\match(G')\cup\hat{M},$$
where $\hat{M}$ is a maximum cardinality matching of agent $i$ on its hidden and unmatched vertices.

The symmetric difference
$$
M\Delta M'=M\cup M'\setminus(M\cap M')
$$
then consists of vertex-disjoint paths (some of which may be cycles) with alternating edges of $M$ and $M'$. For example, consider the two-agent version of $\match$ applied to the graphs $G$ and $G'$ given in Figures~\ref{fig:lb_a} and Figure~\ref{fig:lb_b}. It holds that
$$M=\match(G)=\{(v_2,v_3),(v_4,v_5),(v_6,v_7)\},$$
whereas, say, $M'=\{(v_2,v_3),(v_5,v_6)\}$. Then, $M\Delta M'$ is the single path $\{(v_4,v_5),(v_5,v_6),(v_6,v_7)\}$ where the first and last edge are in $M$ and the middle edge is in $M'$.

In order to simplify notation, we henceforth assume that $M\Delta M'$ consists of just one path. This assumption is made without loss of generality, because we show that each such path satisfies one of the following properties: either $M$ matches at least as many vertices of $V_i$ as $M'$ for every $i\in N$, or one can derive a contradiction to the way $M$ or $M'$ were selected by switching between some (or all) of their edges on the path. Since the contradiction can be derived for each path separately, it follows that the first property holds on every path, that is, the overall utility of agent $i$ for $M$ is at least as large as its utility for $M'$.

If the path in $M\Delta M'$ is a cycle, then this cycle must be of even length, because otherwise there would be a vertex that is incident to two edges of the same matching. It follows that both $M$ and $M'$ match all the vertices on the cycle, hence agent $i$ is indifferent between the two matchings. We may therefore assume that $M\Delta M'$ is not a cycle.

It will prove useful to arbitrarily fix a direction over the (undirected) edges of the single path in $M\Delta M'$. Since the path is not a cycle, this direction pinpoints two specific vertices as the start and the end of the path. We further say that the (directed) edge $(u,v)$ \emph{enters} $V_j$ if $u\notin V_j$ and $v\in V_j$, and \emph{exits} $V_j$ if $u\in V_j$ and $v\notin V_j$.

We consider two cases.
\par\noindent\emph{Case 1:} $|M_{ii}|>|M'_{ii}|$. We claim that
\begin{equation}
\label{eq:diff}
\sum_{j\in N\setminus\{i\}}|M_{ij}| \geq \sum_{j\in N\setminus\{i\}}|M'_{ij}| - 2.
\end{equation}
Since both $M$ and $M'$ are maximum cardinality matchings on $V_j$ for all $j\neq i$, it must hold that every subpath of $M\Delta M'$ on $V_j$ has even length (see Figure~\ref{fig:partition}); otherwise we would have, say, more edges of $M$ than $M'$ on the subpath, and by switching from $M'$ to $M$ on the subpath we would be able to increase the size of $M'$ on $V_j$. This implies that for any $j\in N\setminus\{i\}$, any subpath entering $V_j$ with an edge of $M'$ must exit $V_j$ with an edge of $M$, and any subpath entering $V_j$ with and edge of $M$ must exit $V_j$ with an edge of $M'$.

The next part of the proof is crucial, and uses the main idea behind mechanism $\match$. We argue that it also holds that a subpath exiting $V_i$ with an edge of $M'$ can only enter $V_i$ with an edge of $M$. Assume without loss of generality that $i\in \Pi_1$. By the above argument the subpath enters $V_{j_1}$, ${j_1}\in \Pi_2$, with an edge of $M'$, and therefore exits it with an edge of $M$, entering some $V_{j_2}$ in $\Pi_1$. If $j_2\neq i$, and the subpath exits $V_{j_2}$, then it does so with an edge of $M'$, and by the same arguments returns to the vertex set of an agent in $\Pi_1$ with an edge of $M$. If eventually the subpath enters $V_i$ again, it must be with an edge of $M$. Analogously, if the subpath exits $V_i$ with an edge of $M$, it can only enter $V_i$ with an edge of $M'$. See Figure~\ref{fig:partition} for an illustration.

Now consider $(M\Delta M')\setminus(M_{ii}\cup M'_{ii})$, which again is a collection of vertex-disjoint subpaths. Some start and end in $V_i$, and it follows by the discussion above that such subpaths have exactly one edge in $M_{ij}$ and one edge in $M'_{ik}$, for $k,j\in N\setminus\{i\}$. There can only be one subpath that starts in $V_i$ but does not end in $V_i$, and at most one subpath that ends in $V_i$ but does not start in $V_i$. Equation~\eqref{eq:diff} directly follows.

We now have that
\begin{align*}
u_i(M) &= 2|M_{ii}| + \sum_{j\in N\setminus\{i\}}|M_{ij}|\\
& \geq 2(|M'_{ii}|+1) + \left(\sum_{j\in N\setminus\{i\}}|M'_{ij}|-2\right)\\
& = u_i(M'),
\end{align*}
where the inequality follows from the fact that $|M_{ii}|>|M'_{ii}|$ and from~\eqref{eq:diff}.

\par\noindent\emph{Case 2: $|M_{ii}'|=|{M_{ii}}|$.} Note that it holds that $|M_{jj}|=|M'_{jj}|$ for all $j\in N$, that is, $M\Delta M'$ has to be of even length inside every $V_j$. This includes $M_{ii}$ and $M'_{ii}$, because the total number of internal edges for $i$ is even.  If some subpath of $i$'s internal edges has odd length with more edges from $M$ there must be another subpath with more internal edges from $M'$.  Swapping the edges of $M$ for those of $M'$ in the second subpath results in a matching $M''$ such that $|M_{ii}''| > |M_{ii}|$ contradicting the construction of $M$ to have maximum cardinality on each $V_i$.  It follows that $|M|\geq |M'|$, since $M$ is a maximum cardinality matching under the constraint that it has maximum cardinality inside each $V_i$.

We claim that if $|M|>|M'|$ then $\sum_j |M_{ij}|\geq \sum_j |M'_{ij}|$.  Together with the assumption that $|M_{ii}'|=|{M_{ii}}|$ this implies that agent $i$ cannot benefit. Indeed, in this case $M\Delta M'$ is a path of odd length that starts and ends with an edge of $M$. Recall that every subpath of $M \Delta M'$ consisting of $i$'s internal edges has even length.  This means that when the path enters $V_i$ with an edge of $M'$ it cannot end inside $V_i$, as otherwise it would end with an edge of $M'$. In other words, every time the path enters $V_i$ with an edge of $M'$ it must exit $V_i$ with an edge of $M$. Similarly, every time the path exits $V_i$ with an edge of $M'$ it must have entered $V_i$ with an edge of $M$, otherwise the path must start in $V_i$ with an edge of $M'$. This proves our claim, so $|M|=|M'|$.

Since $|M|=|M'|$ we have that $M\Delta M'$ has even length, and moreover we know it has even length inside each $V_i$. Note that all the vertices on the path are matched under both $M$ and $M'$, except for the start and the end vertices. Hence, if agent $i$ gains from the manipulation, it must be the case (when fixing a specific direction on the edges) that the start vertex is a vertex of $V_i$ and the first edge is an edge of $M'$, whereas the end vertex is in $V_j$, for some $j\in N\setminus\{i\}$, and the last edge is an edge of $M$.

Now, if tie-breaking favors $i$ over $j$, then by switching the edges of $M$ with those of $M'$ we get a matching of equal size that has maximum cardinality on each $V_i$ and is better for $i$, in contradiction to the tie-breaking rule. If tie-breaking favors $j$ over $i$, consider the subpath of $M\Delta M'$ that starts with the last edge that exits $V_i$ and ends with the last edge in $M\Delta M'$. This path must start with an edge of $M'$.  To see why, note that $M \Delta M'$ starts in $V_i$ with an edge of $M'$.  This subpath has even length, so it exits with an edge of $M'$.  By the same argument as in Case 1, the bipartition ensures that, if the path re-enters $V_i$, it does so with an edge from $M$.  Since all subpaths of vertices in $V_i$ are of even length, the path always exits $V_i$ with an edge of $M'$.

By replacing all the edges of $M'$ with the edges of $M$ on this subpath, we can obtain a matching $M''$ that is identical to $M'$ inside $V_i$, has maximum cardinality on $V_k$ for each $k\in N$, is as large as $M'$ overall, and satisfies $u_j(M'')=u_j(M')+1$, $u_i(M')=u_i(M'')-1$, and $u_k(M'')=u_k(M')$ for all $k\in N\setminus\{i,j\}$. By removing the edges of $\hat{M}$ (recall this is the second stage internal matching of $i$) from both $M'$ and $M''$ we get a contradiction to the way the mechanism broke ties when constructing $M'$ (specifically, when constructing $\match(G')$).
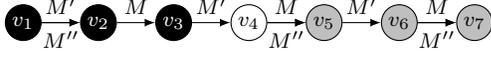
\begin{figure}
\centering
\begin{tikzpicture}
\tikzstyle{graydot}=[circle,draw=black,fill=lightgray,thin,
inner sep=1.5pt,minimum size=1.5mm]
\tikzstyle{blackdot}=[circle,draw=black,fill=black,thin,
inner sep=1.5pt,minimum size=1.5mm]
\tikzstyle{whitedot}=[circle,draw=black,fill=white,thin,
inner sep=1.5pt,minimum size=1.5mm]
\tikzstyle{graydot2}=[circle,draw=black,fill=lightgray,thin,
inner sep=0.3,minimum size=1.5mm]
\tikzstyle{whitedot2}=[circle,draw=black,fill=white,thin,
inner sep=0.3,minimum size=1.5mm]
\tikzstyle{blackdot2}=[circle,draw=black,fill=black,thin,
inner sep=0.3,minimum size=1.5mm]

\node (v1) at (0,0) [blackdot] {\color{white}\small{$v_1$}};
\node (v2) at (1,0) [blackdot] {\color{white}\small{$v_2$}};
\node (v3) at (2,0) [blackdot] {\color{white}\small{$v_3$}};
\node (v4) at (3,0) [whitedot] {\small{$v_4$}};
\node (v5) at (4,0) [graydot] {\small{$v_5$}};
\node (v6) at (5,0) [graydot] {\small{$v_6$}};
\node (v7) at (6,0) [graydot] {\small{$v_7$}};

\path [-latex] (v1) edge node [above] {\small{$M'$}} (v2);
\path [-latex] (v2) edge node [above] {\small{$M$}} (v3);
\path [-latex] (v3) edge node [above] {\small{$M'$}} (v4);
\path [-latex] (v4) edge node [above] {\small{$M$}} (v5);
\path [-latex] (v5) edge node [above] {\small{$M'$}} (v6);
\path [-latex] (v6) edge node [above] {\small{$M$}} (v7);

\path [-latex] (v1) edge node [below] {\small{$M''$}} (v2);
\path [-latex] (v4) edge node [below] {\small{$M''$}} (v5);
\path [-latex] (v6) edge node [below] {\small{$M''$}} (v7);

\end{tikzpicture}
\caption{An illustration of the last argument in Case 2 of the proof of Theorem~\ref{thm:sp} with $i = 3$ and $j = 2$.  The vertices of $V_1$ are white, the vertices of $V_2$ are gray, and the vertices of $V_3$ are black. By switching from $M'$ to $M''$ we increase the utility of agent $2$ and decrease the utility of agent $3$, thereby obtaining a legal matching that contradicts the choice of $M'$.}
\label{fig:partition2}
\end{figure}
See Figure~\ref{fig:partition2} for an illustration.
\end{proof}

We next show that $\match$ can be executed in polynomial time by a reduction to the maximum weighted matching problem (for a polynomial time algorithm
see~\cite{Gabow90}).
\begin{theorem}
$\match$ can be executed in polynomial time.
\end{theorem}
\begin{proof}
Assume without loss of generality that $|E|>1$, and let $\epsilon_i = 1 / |E|^{i+1}$.  We assign weights to edges as follows.  An (internal) edge $(u,v)$ such that $u,v \in V_i$ for some $i\in N$ receives weight $|E| + 3$.  An (external) edge $(u,v)$ such that $u \in V_i$ and $v \in V_j$ with $i\in\Pi_1$ and $j\in\Pi_2$ receives weight $1 + \epsilon_i + \epsilon_j / |E|^{n+1}$.  An (external) edge $(u,v)$ such that $u \in V_i$ and $v \in V_j$ with $i \neq j$ but $i,j \in \Pi_1$ or $i,j \in \Pi_2$ receives weight $0$.

The sum of the weights of all external edges is at most $|E|(1 + 1/|E|^2 + 1/|E|^{n+3}) < |E| + 3$, which is less than the weight of a single internal edge.  Thus a maximum weight matching of this graph maximizes the number of internal edges.  All edges between sets on the same side of the bipartition have weight zero, so no such edges will be included.

To complete the proof we need to verify that the maximum weight matching has maximum cardinality among those with a maximum number of internal edges and no edges across the bipartition, and that ties are broken appropriately.  Each edge across the bipartition has weight at least $1$ and at most $1 + 1/|E|^2 + 1/|E|^{n+3}$. Thus, given two matchings $M$ and $M'$ satisfying the above constraints such that $|M|>|M'|$, the difference in their weights is at least
\begin{align*}
	1 - |M'|(1/|E|^2 + 1/|E|^{n+3}) &\geq  1 - |E|(1/|E|^2 + 1/|E|^{n+3})\\ & =1 - 1/|E| - 1/|E|^{n+2}> 0.
\end{align*}
The maximum weight matching thus has maximum cardinality subject to the constraints.  For tie-breaking, observe that $\epsilon_i\geq|E|\epsilon_j$ if $i<j$, meaning that among agents on the same side of the bipartition those with smaller indices have higher priority.  The factor of $1 / |E|^{n+1}$ finally ensures that agents in $\Pi_1$ have priority over agents in $\Pi_2$.
\end{proof}

Recall that by Theorem~\ref{thm:lb} no deterministic SP mechanism can have an approximation ratio smaller than $2$, even when there are only two agents.  We will see momentarily that $\match$ provides an approximation ratio of $2$ when $N=\{1,2\}$ and $\Pi=(\{1\},\{2\})$, i.e., it is the best possible deterministic SP mechanism for the case of two agents.  Indeed, consider a graph $G$, let $M^*$ be an optimal matching of~$G$, and $M$ the matching returned by $\textsc{Match}_{(\{1\},\{2\})}$.  $M$ is inclusion-maximal. Therefore, for every $(u,v)\in M^*$, either $u$ is matched by $M$ or $v$ is matched by $M$.  We conclude that $|M|\geq |M^*|/2$. Strategyproofness is obtained from Theorem~\ref{thm:sp}.
\begin{corollary}
\label{cor}
Let $N=\{1,2\}$. Then, $\textsc{Match}_{(\{1\},\{2\})}$ is SP and provides a $2$-approximation with respect to social welfare.
\end{corollary}

Unfortunately, when $n\geq 3$, $\match$ does not provide a finite approximation ratio for any fixed bipartition. To see this, let $\Pi=(\Pi_1,\Pi_2)$ be a bipartition of the set of agents. Then there must be two distinct agents $i,j\in N$ such that $i,j\in \Pi_l$ for some $l\in \{1,2\}$. Now consider a graph where the only edge is an external edge between $V_i$ and $V_j$; given this graph $\match$ returns an empty matching, whereas the optimum is a matching of cardinality $1$.

We believe that in general deterministic SP mechanisms can only provide a bad approximation ratio, even for the case of three agents. The following conjecture makes this precise.
\begin{conjecture}
\label{conj:det_lb}
If there are more than two agents, no deterministic SP mechanism can provide an $\alpha$-approximation with respect to social welfare for any constant~$\alpha$.
\end{conjecture}

\section{Randomized mechanisms}
\label{sec:rand}

We have seen above that $\match$ does not provide a bounded approximation ratio for any fixed bipartition $\Pi$. The natural next step is to choose the bipartition uniformly at random. This leads to the eponymous \mixnmatch mechanism.

\par\medskip
\noindent\mixnmatch
\begin{enumerate}
\item \textbf{Mix:} Construct a random bipartition $\Pi=(\Pi_1,\Pi_2)$ of the agents by independently flipping a fair coin for each agent to determine whether the agent is in $\Pi_1$ or in $\Pi_2$.
\item \textbf{Match:} Apply $\match$ to the given graph, where $\Pi$ is the bipartition constructed in step 1.
\end{enumerate}

It immediately follows from Theorem~\ref{thm:sp} that \mixnmatch is SP, and in fact in a stronger sense than the one defined in Section~\ref{sec:prem}, namely \emph{universal strategyproofness}. A randomized mechanism is called universally SP if agents cannot gain by lying regardless of the random choices made by the  mechanism, i.e., if the mechanism is a distribution over SP deterministic mechanisms.

A na\"ive analysis of \mixnmatch would yield a rather unimpressive approximation ratio. Indeed, the reason why $\textsc{Match}_{(\{1\},\{2\})}$ does not provide a better approximation ratio than two is that it may have to sacrifice two external edges for one internal edge. The fact that \mixnmatch will not be able to match many of the edges in the graph because they are not between the two elements of the constructed bipartition would seem to cause the approximation ratio to deteriorate further. Fortunately, these two problems effectively cancel out: sacrificing two external edges for an internal edge is less of a problem when each of those external edges is allowed to be part of the matching for only half of the bipartitions.  Formally, we prove the following result.
\begin{theorem}
\label{thm:approx}
For any number of agents, \mixnmatch is (universally) SP and provides a $2$-approximation with respect to social welfare.
\end{theorem}
\begin{proof}
We prove the theorem by taking a maximum cardinality matching $M^*$ and constructing a matching $M'$ that, when restricted according to a random bipartition (by removing edges between agents on the same side of the bipartition), has at least half the size of $M^*$ in expectation.  The matching produced by $\match$ then always is at least as large as $M'$ restricted according to $\Pi$.

Consider a graph $G$, and let $M^*$ be a maximum cardinality matching of $G$. For each $i\in N$ let $M_i^{**}$ be a maximum cardinality matching on $V_i$, and let $M^{**} = \bigcup_{i\in N} M_i^{**}$.

We construct a matching $M'$ by considering the symmetric difference $M^* \Delta M^{**}$.  As in the proof of Theorem~\ref{thm:sp}, it consists of a set of paths with alternating edges of $M^*$ and $M^{**}$.  For each path, if there are more internal edges among the edges from $M^{**}$, we put those edges in $M'$.  Otherwise, we put the edges from $M^*$ in $M'$.

Since $M^{**}$ has maximum cardinality on each $V_i$ and $M'$ has the same number of internal edges from each path as $M^{**}$, $M'$ has maximum cardinality on each $V_i$.  Furthermore, since $M^*$ is a maximum cardinality matching, each path has either the same number of edges from $M^*$ and $M^{**}$ or one extra edge from $M^*$.  All external edges on the path are from $M^*$, so if the edges from $M^{**}$ are taken for $M'$ then the number of internal edges gained relative to $M^*$ is at least the number of external edges lost minus one.  In the worst case $M'$ has two fewer external edges for each extra internal edge relative to $M^*$.
Thus $M'$ satisfies
$$
\sum_{i\in N} (|M'_{ii}|-|M^*_{ii}|)\geq  \frac{1}{2}\sum_{i\in N}\sum_{j>i}(|M^*_{ij}|-|M'_{ij}|),
$$
where we sum over $j>i$ so as not to count the same edges twice. Rearranging, we get
\begin{multline}
\label{eq:rearrange}
\sum_{i\in N} |M'_{ii}| +  \frac{1}{2}\sum_{i\in N}\sum_{j>i}|M'_{ij}|\geq \\ \sum_{i\in N} |M^*_{ii}| +  \frac{1}{2}\sum_{i\in N}\sum_{j>i}|M^*_{ij}|.
\end{multline}

Now let $M^\Pi$ be the matching produced by $\match$ for the fixed bipartition $\Pi$. Since $M^\Pi$ has maximum cardinality under the constraints, we have
\begin{align*}
|M^\Pi|&=\sum_{i\in N} |M^\Pi_{ii}| + \sum_{i\in \Pi_1}\sum_{j\in \Pi_2}|M^\Pi_{ij}|\\
& \geq \sum_{i\in N} |M'_{ii}| + \sum_{i\in \Pi_1}\sum_{j\in \Pi_2}|M'_{ij}|.
\end{align*}

Since each pair of agents appears on opposite sides in exactly half of the bipartitions, the expected size of the matching produced by \mixnmatch is
\begin{align*}
\sum_\Pi \left(\frac{1}{2^n}\cdot |M^\Pi|\right)&\geq \sum_{i\in N} |M'_{ii}| +  \frac{1}{2}\sum_{i\in N}\sum_{j>i}|M'_{ij}|\\
&\geq \sum_{i\in N} |M^*_{ii}| +  \frac{1}{2}\sum_{i\in N}\sum_{j>i}|M^*_{ij}|\\
& \geq \frac{1}{2}\cdot |M^*|,
\end{align*}
where the second inequality follows from~\eqref{eq:rearrange}.
\end{proof}

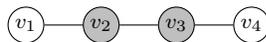
\begin{figure}
\centering
\begin{tikzpicture}
\tikzstyle{graydot}=[circle,draw=black,fill=lightgray,thin,
inner sep=1.5pt,minimum size=1.5mm]
\tikzstyle{blackdot}=[circle,draw=black,fill=black,thin,
inner sep=1.5pt,minimum size=1.5mm]
\tikzstyle{whitedot}=[circle,draw=black,fill=white,thin,
inner sep=1.5pt,minimum size=1.5mm]
\tikzstyle{graydot2}=[circle,draw=black,fill=lightgray,thin,
inner sep=0.3,minimum size=1.5mm]
\tikzstyle{dasheddot}=[circle,dashed,draw=black,fill=white,thin,
inner sep1.5pt,minimum size=1.5mm]

\node (v1) at (0,0) [whitedot] {\small{$v_1$}};
\node (v2) at (1,0) [graydot] {\small{$v_2$}};
\node (v3) at (2,0) [graydot] {\small{$v_3$}};
\node (v4) at (3,0) [whitedot] {\small{$v_4$}};

\draw (v1.east) -- (v2.west);
\draw (v2.east) -- (v3.west);
\draw (v3.east) -- (v4.west);

\end{tikzpicture}
\caption{Graph illustrating that Mix-and-Match cannot provide an approximation ration smaller than two.  $V_1$ is shown in white, $V_2$ is shown in gray.  Mix-and-Match returns the matching $(v_2,v_3)$.}
\label{fig:tight}
\end{figure}
The graph in Figure~\ref{fig:tight} shows that the analysis of \mixnmatch is tight even for $n=2$.  Still one might hope to do better, given that Theorem~\ref{thm:lb} only provides a randomized lower bound of $4/3$. Consider the following randomized mechanism for the case of two agents.

\par\medskip
\noindent\textsc{Flip-and-Match}
\begin{enumerate}
\item Given a graph $G$, flip a fair coin.
\item If the outcome is heads, return $\textsc{Match}_{(\{1\},\{2\})}(G)$.
\item If the outcome is tails, choose a maximum cardinality matching, breaking ties in favor of a matching that maximizes the total number of internal edges and then arbitrarily.
\end{enumerate}

With probability $1/2$ this mechanism uses $\textsc{Match}_{(\{1\},\{2\})}$ and returns a matching of cardinality at least half the optimum. With probability $1/2$ the mechanism returns a maximum cardinality matching. Hence, the expected cardinality of the matching returned by the mechanism is at least~$3/4$ of the cardinality of an optimal one.  We conclude that \textsc{Flip-and-Match} provides an approximation ratio of $4/3$. Whether it is SP however remains an open question.
\begin{conjecture}
\label{conj:rand_ub}
\textsc{Flip-and-Match} is SP (in expectation).
\end{conjecture}

We justify and discuss this conjecture in Appendix~\ref{app:rand_ub}. Note that \textsc{Flip-and-Match} is similar to Mechanism~1 of Procaccia and Tennenholtz~\cite{PT09}, in the sense that it selects an optimal solution with probability $1/2$ and an SP solution with probability $1/2$.


\section*{Discussion}

Our results have shown that \mixnmatch is near optimal.  Theorem~\ref{thm:approx} shows that it is a 2-approximation, which according to Theorem~\ref{thm:lb} is the lower bound for deterministic mechanisms and near the lower bound for randomized mechanisms.  However, a 2-approximation is a 50\% efficiency reduction and may not be acceptable in practice for kidney exchanges.  The bipartition that \mixnmatch uses is also somewhat problematic: it may be hard to convince hospitals that they best serve their patients by refusing to match them with patients from half the other hospitals.  Our results provide several insights into these issues.

Our results deal with worst case guarantees.  If these are essential, then our lower bounds show
that nothing better can be done so strategyproofness will have to be
relaxed.  An alternative is to consider less strict notions of incentive compatibility.  This is the route taken in \cite{AGR09}.

While the worst case performance of \mixnmatch might be problematic, the average case performance may be significantly better.  In particular, much of the efficiency loss comes from being unable to match patients of hospitals assigned to same side of the bipartition.  However, in a large market with many hospitals and many potential matches for each patient this may not be a significant restriction. Our guess based on related work is that the loss probably would be
quite small (on the order of a few percent).  However, like many
average-case questions, the results are sensitive to what is chosen as
the underlying distribution.  This is an interesting direction for future research.

Alternatively, an algorithm like the naive 3-agent algorithm might be ``close
enough to strategy-proof'' in practice.  If examples like that of Figure~\ref{fig:counter} are
rare, then there might be few enough opportunities for manipulation
that the results will be tolerable and the use of the bipartition could
be eliminated.  Note however that such an algorithm in still only
2-approximate in the worst case.

Another direction would be to try and characterize the space of strategyproof mechanisms, although our results
suggest that there probably is not a simple characterization.
The $\match$ framework does capture a number of obvious strategyproof mechanisms
(for example, only taking internal edges corresponds to a bipartition
with all agents on the same side).  However, a mechanism like
selecting 2 agents and then running the 2 agent mechanism on those
agents does not correspond to any bipartition.  If our conjecture that
\textsc{Flip-and-Match} is strategyproof is correct, that is another mechanism
that does not fit in this framework.  While all of these examples are
"$\match$ like", we also know of at least one (relatively complex) mechanism that looks quite different.

\section*{Acknowledgments}

We thank Moshe Tennenholtz for his important contributions to this paper. We also thank David Parkes for helpful discussions and for his detailed comments on a draft of this paper.
This material is based on work supported by the Deutsche Forschungsgemeinschaft under grant FI~1664/1-1.

\bibliographystyle{plain}
\bibliography{abb,ultimate}

\begin{thebibliography}{10}

\bibitem{ABS07}
D.~Abraham, A.~Blum, and T.~Sandholm.
\newblock Clearing algorithms for barter exchange markets: Enabling nationwide
  kidney exchanges.
\newblock In {\em Proceedings of the 8th ACM Conference on Electronic Commerce
  (EC)}, pages 295--304, 2007.

\bibitem{AFPT09}
N.~Alon, M.~Feldman, A.~D. Procaccia, and M.~Tennenholtz.
\newblock Strategyproof approximation mechanisms for location on networks.
\newblock Manuscript, 2010.

\bibitem{AFPT09b}
N.~Alon, F.~Fischer, A.~D. Procaccia, and M.~Tennenholtz.
\newblock Sum of us: Strategyproof selection from the selectors.
\newblock Manuscript, 2010.

\bibitem{AGR09}
I.~Ashlagi, D.~Gilchrist, and A.E. Roth.
\newblock New incentives in large scale kidney exchange.
\newblock Manuscript, 2010.

\bibitem{BRM09}
P.~Bir\'o, D.~F. Manlove, and R.~Rizzi.
\newblock Maximum weight cycle packing in directed graphs, with application to
  kidney exchange programs.
\newblock {\em Discrete Mathematics, Algorithms and Applications},
  1(4):499--517, 2009.

\bibitem{CKV09}
G.~Christodoulou, E.~Koutsoupias, and A.~Vidali.
\newblock A lower bound for scheduling mechanisms.
\newblock {\em Algorithmica}, 55(4):729--740, 2009.

\bibitem{DFP10}
O.~Dekel, F.~Fischer, and A.~D. Procaccia.
\newblock Incentive compatible regression learning.
\newblock {\em Journal of Computer and System Sciences}.
\newblock To Appear.

\bibitem{DDDR08}
P.~Dhangwatnotai, S.~Dobzinski, S.~Dughmi, and T.~Roughgarden.
\newblock Truthful approximation schemes for single-parameter agents.
\newblock In {\em Proceedings of the 49th Symposium on Foundations of Computer
  Science (FOCS)}, pages 15--24, 2008.

\bibitem{DG10}
S.~Dughmi and A.~Ghosh.
\newblock Truthful assignment without money.
\newblock In {\em Proceedings of the 11th ACM Conference on Electronic Commerce
  (EC)}, 2010.
\newblock To Appear.

\bibitem{Gabow90}
H.~N. Gabow.
\newblock Data structures for weighted matching and nearest common ancestors
  with linking.
\newblock In {\em Proceedings of the 1st Annual ACM-SIAM Symposium on Discrete
  Algorithms (SODA)}, pages 434--443, 1990.

\bibitem{GC10}
M.~Guo and V.~Conitzer.
\newblock Strategy-proof allocation of multiple items without payments or
  priors.
\newblock In {\em Proceedings of the 9th International Joint Conference on
  Autonomous Agents and Multi-Agent Systems (AAMAS)}, 2010.
\newblock To Appear.

\bibitem{LS05}
R.~Lavi and C.~Swami.
\newblock Truthful and near-optimal mechanism design via linear programming.
\newblock In {\em Proceedings of the 46th Symposium on Foundations of Computer
  Science (FOCS)}, pages 595--604, 2005.

\bibitem{LOS02}
D.~Lehmann, L.~I. O'Callaghan, and Y.~Shoham.
\newblock Truth revelation in rapid, approximately efficient combinatorial
  auctions.
\newblock {\em Journal of the ACM}, 49(5):577--602, 2002.

\bibitem{LWZ09}
P.~Lu, Y.~Wang, and Y.~Zhou.
\newblock Tighter bounds for facility games.
\newblock In {\em Proceedings of the 5th International Workshop on Internet and
  Network Economics (WINE)}, pages 137--148, 2009.

\bibitem{MPR08b}
R.~Meir, A.~D. Procaccia, and J.~S. Rosenschein.
\newblock Strategyproof classification under constant hypotheses: A tale of two
  functions.
\newblock In {\em Proceedings of the 23rd AAAI Conference on Artificial
  Intelligence (AAAI)}, pages 126--131, 2008.

\bibitem{MPR09}
R.~Meir, A.~D. Procaccia, and J.~S. Rosenschein.
\newblock Strategyproof classification with shared inputs.
\newblock In {\em Proceedings of the 21st International Joint Conference on
  Artificial Intelligence (IJCAI)}, pages 220--225, 2009.

\bibitem{NR01}
N.~Nisan and A.~Ronen.
\newblock Algorithmic mechanism design.
\newblock {\em Games and Economic Behavior}, 35(1--2):166--196, 2001.

\bibitem{PT09}
A.~D. Procaccia and M.~Tennenholtz.
\newblock Approximate mechanism design without money.
\newblock In {\em Proceedings of the 10th ACM Conference on Electronic Commerce
  (EC)}, pages 177--186, 2009.

\bibitem{RothKidneyQJE}
A.~E. Roth, T.~S\"onmez, and M.~U. \"Unver.
\newblock {Kidney exchange}.
\newblock {\em Quarterly Journal of Economics}, 119:457--488, 2004.

\bibitem{RothKidneyAERPP}
A.~E. Roth, T.~S\"onmez, and M.~U. \"Unver.
\newblock {A kidney exchange clearinghouse in New England}.
\newblock {\em American Economic Review: Papers and Proceedings},
  95(2):376--380, 2005.

\bibitem{RothKidneyJET}
A.~E. Roth, T.~S\"onmez, and M.~U. \"Unver.
\newblock {Pairwise kidney exchange}.
\newblock {\em Journal of Economic Theory}, 125:151--188, 2005.

\bibitem{RSMunpublished}
A.~E. Roth, T.~S{\"o}nmez, and M.~U. {\"U}nver.
\newblock {Transplant Center Incentives in Kidney Exchange}.
\newblock Unpublished, 2005.

\bibitem{RothKidneyAER}
A.~E. Roth, T.~S\"onmez, and M.~U. \"Unver.
\newblock {Efficient kidney exchange: coincidence of wants in markets with
  compatibility-based preferences}.
\newblock {\em American Economic Review}, 97:828--851, 2007.

\end{thebibliography}

\appendix

\section{A discussion of Flip-and-Match}
\label{app:rand_ub}

In order to gain some intuition, let us apply \textsc{Flip-and-Match} to the example graph in Figure~\ref{fig:lb_a}. For this graph, $\textsc{Match}_{(\{1\},\{2\})}$ returns the matching
$$
\{(v_2,v_3),(v_4,v_5),(v_6,v_7)\},
$$
which is also the unique maximum cardinality matching that maximizes the number of internal edges. Hence, the utility of agent $1$ is $3$. When agent $1$ hides $v_5$ and $v_6$ (Figure~\ref{fig:lb_b}, $\textsc{Match}_{(\{1\},\{2\})}$ returns the matching $\{(v_2,v_3)\}$, whereas the unique maximum cardinality matching is $\{(v_1,v_2),(v_3,v_4)\}$. The expected utility of agent $1$, also taking into account its internal matching on $(v_5,v_6)$, is $3$.

While in both cases agent $1$'s expected utility is 3, this happens in different ways.  If agent $1$ does not deviate his utility is 3 regardless of which algorithm is used.  However, if he does deviate he gains when the maximum cardinality matching is used, but he loses exactly the same amount when $\textsc{Match}_{(\{1\},\{2\})}$ is used.  Thus, randomizing uniformly prevents an incentive to misreport.  This phenomenon occurs in all examples we have considered.

We know from Theorem~\ref{thm:sp} that an agent can never gain under $\textsc{Match}_{(\{1\},\{2\})}$ by hiding some of its vertices.  It might however gain under the maximum cardinality matching.  In order to prove that \textsc{Flip-and-Match} is SP one would have to show that any gain under a maximum cardinality matching would imply a loss by at least the same amount under $\textsc{Match}_{(\{1\},\{2\})}$.  However, proving such a claim requires considering four different matchings, which causes the techniques used in the proof of Theorem~\ref{thm:sp} to fall short.

The following observation is however somewhat encouraging. \emph{A priori} it seems that \textsc{Flip-and-Match} cannot be SP, since on the same graph there may be several maximum cardinality matchings that maximize the number of internal edges, where some are better for agent $1$ and some are better for agent $2$. Since ties are broken arbitrarily, by hiding some disconnected vertex an agent could in principle cause the mechanism to switch between two such matchings. The following surprising lemma rules out this potential difficulty.
\begin{lemma}\label{lem:lemma}
Let $N=\{1,2\}$. Then for any two maximum cardinality matchings $M$ and $M'$ such that $|M_{11}|+|M_{22}|=|M'_{11}|+|M'_{22}|$ it holds that $u_1(M)=u_1(M')$ and $u_2(M)=u_2(M')$.
\end{lemma}
\begin{proof}[sketch]
Let $M$ and $M'$ be two such matchings, and assume for contradiction that $u_1(M)\neq u_1(M')$. As in the proof of Theorem~\ref{thm:sp}, consider the symmetric difference $M\Delta M'$, and assume without loss of generality that it consists of a single path with alternating edges of $M$ and $M'$. Since $|M|=|M'|$, the path must be of even length. Therefore, since agents $1$ and $2$ have different utilities, the path must have one end in $V_1$ and the other end in $V_2$. This implies that the number of edges between $V_1$ and $V_2$ on the path, that is, the number $|M_{12}|+|M'_{12}|$ is odd.

On the other hand,
$$
|M_{11}|+|M_{22}| + |M_{12}| = |M|=|M'| = |M'_{11}|+|M'_{22}| + |M'_{12}|
$$
and $|M_{11}|+|M_{22}|=|M'_{11}|+|M'_{22}|$, so we have that $|M_{12}|=|M'_{12}|$. In particular $|M_{12}|+|M'_{12}|$ is even. We have reached a contradiction.
\end{proof}

\begin{figure}
\centering
\begin{tikzpicture}
\tikzstyle{graydot}=[circle,draw=black,fill=lightgray,thin,
inner sep=1.5pt,minimum size=1.5mm]
\tikzstyle{blackdot}=[circle,draw=black,fill=black,thin,
inner sep=1.5pt,minimum size=1.5mm]
\tikzstyle{whitedot}=[circle,draw=black,fill=white,thin,
inner sep=1.5pt,minimum size=1.5mm]
\tikzstyle{graydot2}=[circle,draw=black,fill=lightgray,thin,
inner sep=0.3,minimum size=1.5mm]
\tikzstyle{dasheddot}=[circle,dashed,draw=black,fill=white,thin,
inner sep1.5pt,minimum size=1.5mm]

\node (v1) at (0,0) [whitedot] {\small{$v_1$}};
\node (v2) at (1,0) [whitedot] {\small{$v_2$}};
\node (v3) at (2,0) [whitedot] {\small{$v_3$}};
\node (v4) at (3,0) [blackdot] {\color{white}\small{$v_4$}};
\node (v5) at (4,0) [graydot] {\small{$v_5$}};
\node (v6) at (5,0) [graydot] {\small{$v_6$}};
\node (v7) at (6,0) [graydot] {\small{$v_7$}};

\draw (v1.east) -- (v2.west);
\draw (v2.east) -- (v3.west);
\draw (v3.east) -- (v4.west);
\draw (v4.east) -- (v5.west);
\draw (v5.east) -- (v6.west);
\draw (v6.east) -- (v7.west);

\end{tikzpicture}
\caption{Graph illustrating that Lemma~\ref{lem:lemma} cannot be generalized to the case of more than two agents. As usual, vertices of $V_1$ are shown in white, vertices of $V_2$ in gray, and vertices of $V_3$ in black.}
\label{fig:lemma}
\end{figure}
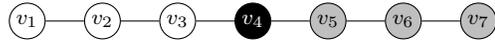
The generalization of this lemma to more than two agents turns out to be false. For $n=3$, consider the graph in Figure~\ref{fig:lemma}.  Both
$$
M=\{(v_1,v_2),(v_3,v_4),(v_5,v_6)\}
$$
and
$$
M'=\{(v_2,v_3),(v_4,v_5),(v_6,v_7\}
$$
are maximum cardinality matchings of this graph that maximize the number of internal edges, but $u_1(M)\neq u_1(M')$ and $u_2(M)\neq u_2(M')$.

\end{document}